\documentclass[preprint,numbers,1p]{elsarticle}

\usepackage{amsmath}
\usepackage{amssymb}
\usepackage{graphicx}
\usepackage{hyperref}
\usepackage{xcolor}
\usepackage{booktabs}
\usepackage{siunitx}
\usepackage{listings}
\usepackage{geometry}

\begin{document}

\begin{frontmatter}

\title{Design optimization of the inter-electrode structures in DC-RSD}

\author[INFN-TO]{N. Cartiglia\corref{mycorrespondingauthor}}
\cortext[mycorrespondingauthor]{Corresponding author}
\ead{cartiglia@to.infn.it}
\author[UPO,INFN-TO]{R. Arcidiacono}
\author[CERN]{G. Bardelli}
\author[FBK,TIFPA]{M. Boscardin}
\author[INFN-FI]{A. Cassese}
\author[FBK,TIFPA]{M. Centis Vignali}
\author[INFN-PG]{T. Croci} 
\author[Unito]{M. Durando}
\author[INFN-TO]{\\ M. Ferrero}
\author[CERN,UniPG]{A. Fondacci}
\author[Unito]{S. Gaiero}
\author[FBK,TIFPA]{O. Hammad Ali}
\author[Unito]{L. Lanteri}
\author[CERN]{A. Losana}
\author[CERN,INFN-TO]{L. Menzio}
\author[INFN-PG]{A. Morozzi}
\author[CNR,INFN-PG]{F. Moscatelli}
\author[UniPG,INFN-PG]{D. Passeri}
\author[FBK,TIFPA]{G.Paternoster}
\author[INFN-FI]{G. Sguazzoni}
\author[INFN-TO]{F. Siviero}
\author[Unito,INFN-TO]{V. Sola}
\author[INFN-FI]{L. Viliani}

\address[UPO]{Universit\`a degli Studi del Piemonte Orientale, largo G. Donegani 2/3,28100 Novara, Italy}
\address[INFN-TO]{Istituto Nazionale di Fisica Nucleare, Sezione di Torino, via P. Giuria 1, 10125 Torino, Italy}
\address[INFN-FI]{Istituto Nazionale di Fisica Nucleare, Sezione di Firenze, Via Bruno Rossi 1, 50019 Sesto Fiorentino (FI), Italy}
\address[Unito]{Universit\`a degli Studi di Torino, via P. Giuria 1, 10125 Torino, Italy}
\address[CNR]{CNR-IOM, Sede secondaria di Perugia, via A. Pascoli 23c, 06123 Perugia, Italy}
\address[INFN-PG]{Istituto Nazionale di Fisica Nucleare, Sezione di Perugia, via A. Pascoli 23c, 06123 Perugia, Italy}
\address[FBK]{Fondazione Bruno Kessler, via Sommarive 18, 38123 Povo (TN), Italy}
\address[TIFPA]{TIFPA-INFN, via Sommarive 18, 38123 Povo (TN), Italy}
\address[UniPG]{Universit\`a degli Studi di Perugia, via G. Duranti 93, 06125 Perugia, Italy}
\address[CERN]{1211 Geneva 23, Switzerland}

\begin{abstract}
DC-coupled Resistive Silicon Detectors (RSD) share charge among electrodes through a
resistive n$^+$ surface layer, and the choice of inter-electrode structure --- a trench, a
resistive strip, or no structure at all --- directly determines the achievable position
resolution and its uniformity. This paper addresses the question of which structure best
optimises the spatial resolution. To answer it quantitatively,
\texttt{DCRSD\_SurfaceSimulator} was developed, a ROOT-based tool that models the resistive surface as a
two-dimensional Kirchhoff resistive network and solves for
charge sharing as a function of hit position, surface resistivity, and inter-electrode
structure. The simulator is validated against test-beam data from FBK DC-RSD sensors with
300, 500, and 1000~\si{\micro\meter} pixel pitch, reproducing the position resolution, its
spatial uniformity, and the signal-sharing templates correctly. Using the
validated simulator, trenches, resistive strips, and no containment are then compared on
position resolution and its spatial uniformity, inter-electrode resistance, and signal
leakage to outer pixels. Resistive strips are found to outperform trenches in resolution by
approximately 25\% for all tested strip resistances (500--2000~$\Omega$) and input impedances
(0--100~$\Omega$), at a surface resistivity of 1000~$\Omega/\square$, and trace the origin of this difference to
the spatial gradient of the charge-sharing fraction: trenches produce a markedly weaker and
less uniform gradient, particularly away from the pixel centre, than either a resistive
strip or a bare resistive surface.
\end{abstract}


\end{frontmatter}

\section{Introduction}
\label{sec:intro}

Resistive Silicon Detectors (RSD) reconstruct the hit position by sharing the collected
charge among several electrodes rather than confining it to a single one
\cite{Menzio2022}. The
fraction of charge reaching each electrode, and how quickly that fraction changes with
position, is affected by the structure placed between electrodes: a physical trench, a
resistive strip, or no structure at all. This raises a direct design question: \emph{which
inter-electrode structure optimises the spatial resolution?} Current DC-RSD sensors, which
use trench isolation, achieve excellent average position resolution but show significant
non-uniformity over the pixel area, making the question a practical one for future
productions.

Answering it quantitatively requires computing the effective impedance between an arbitrary
hit position and every electrode, for each candidate structure — a calculation that, as
shown below, is not tractable analytically. \texttt{DCRSD\_SurfaceSimulator} was therefore
developed, a numerical tool that discretises the resistive surface as
a two-dimensional resistor network and solves the resulting linear system for any injection
position and inter-electrode structure. This study complements ongoing TCAD-based
investigations of DC-RSD design~\cite{Fondacci2025TCAD}, which resolve the detailed
device physics at the expense of computation time, by providing a fast resistor-network
model tailored specifically to the inter-electrode charge-sharing problem, enabling the
broad parameter scans needed to answer the design question above. The rest of this paper
is organised around the three steps of the study: Sections~\ref{sec:sensors}--\ref{sec:reco} describe the simulation
model, Section~\ref{sec:validation} validates it against test-beam data, and
Section~\ref{sec:optimisation} uses the validated model to answer the design question by
comparing trenches, resistive strips, and no containment.

\begin{figure}[!ht]
  \vspace{.2cm}
  \centering
  \includegraphics[width=.95\linewidth]{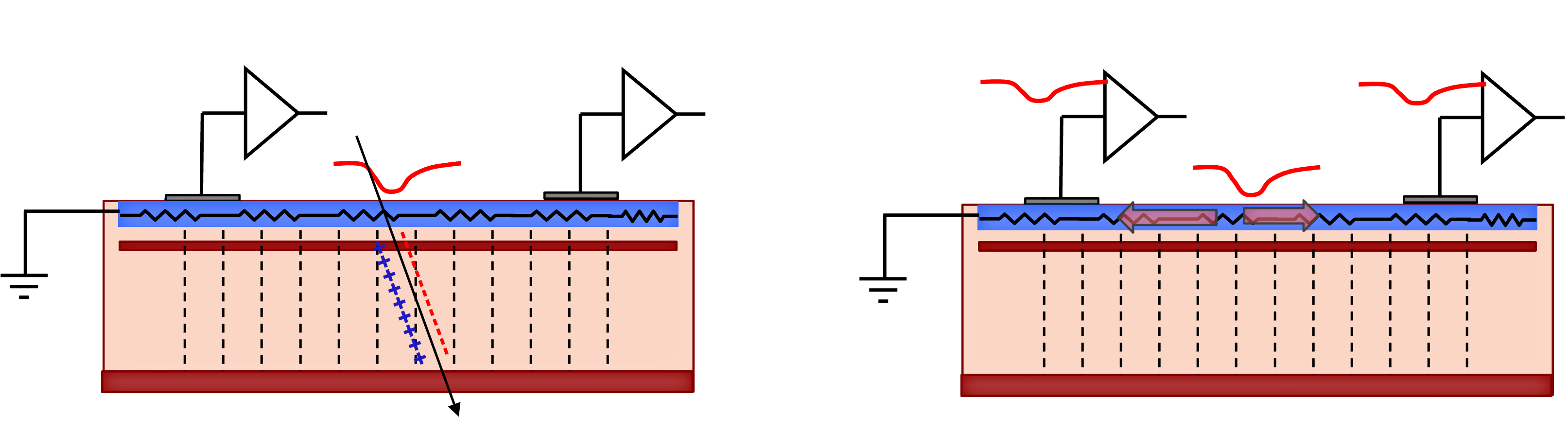}
  \caption{Signal formation in resistive silicon detectors (RSD)
involves two distinct physical processes: (i)  charge generation, drift, and
amplification in the silicon bulk, (ii)  charge redistribution on the resistive
n$^+$ surface layer among the electrodes}
 \label{fig:WF2_SS}
\end{figure}

Signal formation in RSD involves two distinct physical processes. The first — charge
generation, drift, and amplification in the silicon bulk~\cite{Cartiglia2026LGAD} — is well understood and modelled
by several simulators, including Weightfield2~\cite{Cenna2015}. The second — charge
redistribution on the resistive n$^+$ surface layer among the electrodes — is the
subject of this work. Fig.~\ref{fig:WF2_SS} sketches these two processes.

In an RSD, the generated charge is deposited on the resistive surface and spreads laterally
to the electrodes. The signal seen by each electrode is determined by the relative impedances
of the paths connecting the injection point to the electrodes, following a current-divider
law:
\begin{equation}
  S_i \propto \frac{1/Z_i}{\displaystyle\sum_{j=1}^{N} 1/Z_j},
\end{equation}
where $Z_i$ is the effective impedance between the hit position and electrode $i$. While this
picture is formally correct, $Z_i$ depends on position and on the inter-electrode structure
in a way that has no closed-form solution for an arbitrary geometry, which is why a numerical
approach is required.

\section{Experimental sensors}
\label{sec:sensors}

The simulation is validated using three test-beam data sets (TB7, TB8, TB9) collected at the DESY test-beam facility
\cite{DESYTestbeam} from the first FBK DC-RSD production \cite{CentisVignali2025}; for a
broader overview of the status and perspectives of this technology, see \cite{Arcidiacono2026}.
Three pixel geometries, 300, 500, and 1000~\si{\micro\meter} pitch,
and two surface resistivities, $\rho_s = 800$~$\Omega/\square$ and  $\rho_s = 1500$~$\Omega/\square$   are used, all with
trench isolation between pixels. The hit position
reference is provided by a silicon telescope with track resolution of about 
$7-8$~\si{\micro\meter}. The measured position resolution of DC-RSD sensors can be parametrized as the sum of a jitter component, decreasing with amplitude, and a constant term, due to the experiment apparatus.
\begin{equation}
  \sigma_x = p\,\sqrt{\left(\frac{k_1}{\mathrm{MPV}}\right)^2 + k_2^2},
  \label{eq:res_trend}
\end{equation}
where $p$ is the pixel pitch, MPV is the most-probable signal amplitude, and the fitted
constants are $k_1 = 2.5$~mV and $k_2 = 0.03$, the latter being the relative constant term~\cite{Cassese2026}.

\section{Simulation model}
\label{sec:model}

The simulation models a $4\times4$ array of RSD electrodes as a two-dimensional resistive network
on a uniform rectangular grid. Each grid node represents a small patch of the detector surface.
An ionising particle deposits charge at a given position; the charge then spreads through the
resistive layer and is collected by the electrodes. The collected signals are used to
reconstruct the hit position.

\begin{figure}[h]
\centering
 \includegraphics[width=\linewidth]{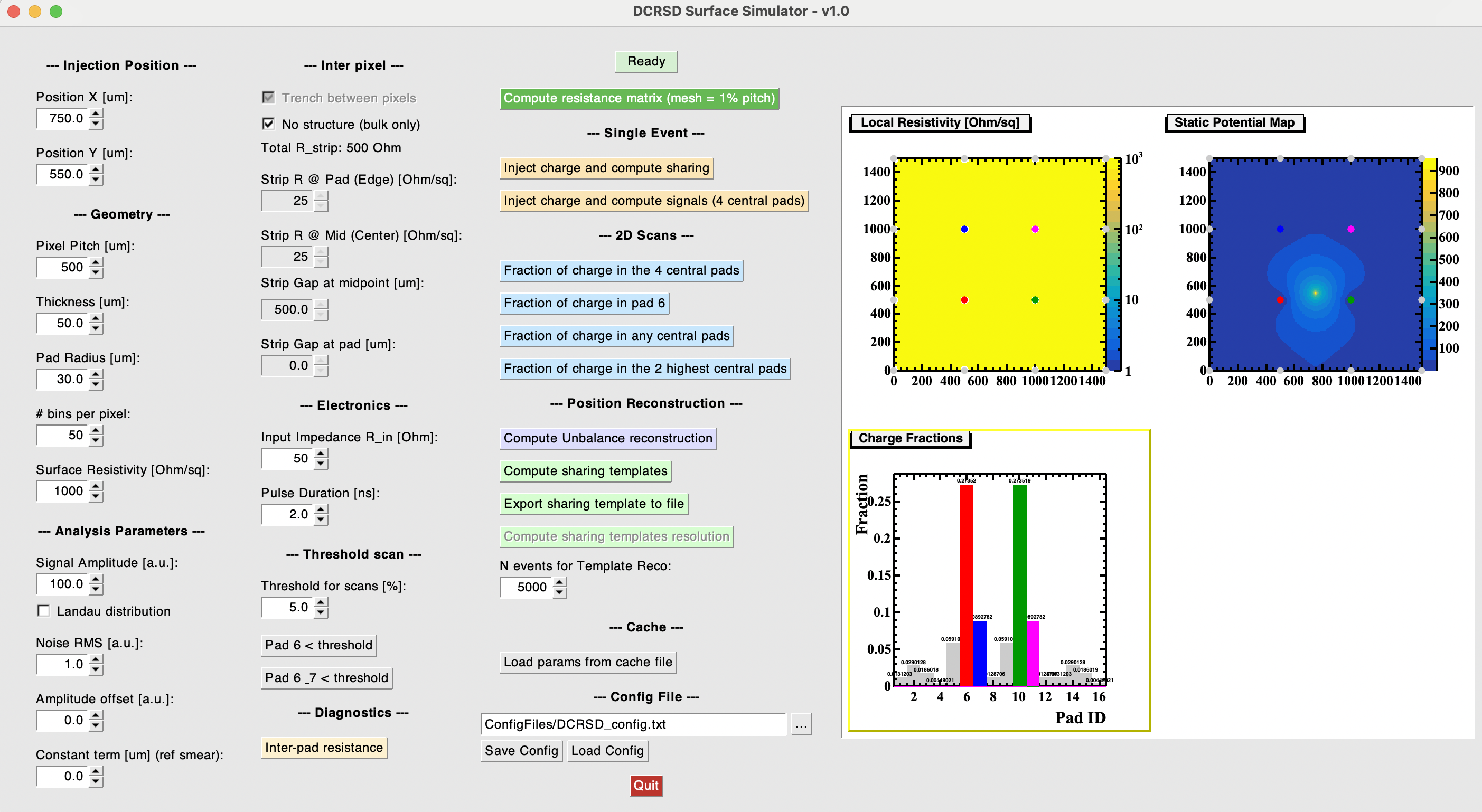}
\caption{Graphical interface of \texttt{DCRSD\_SurfaceSimulator.c}. Left: injection
position, pixel geometry, inter-pixel structure, electronics, and analysis parameters.
Centre: control buttons for single-event, 2D-scan, position-reconstruction, caching, and
configuration-file operations. Right: live plots of the local resistivity map, the static
potential map, and the resulting charge fraction on each electrode.}
\label{fig:gui}
\end{figure}

This model is implemented in the ROOT macro \texttt{DCRSD\_SurfaceSimulator.c}. Its
graphical interface, shown in Fig.~\ref{fig:gui}, exposes all simulation parameters listed
in Table~\ref{tab:params} (injection position, pixel geometry, inter-pixel structure, and
electronics parameters) and gives direct access to the charge-injection, resistance-matrix,
charge-sharing-template, and position-reconstruction tools described in the following
sections, together with live views of the local resistivity map, the static potential map,
and the resulting charge fractions on each electrode.

The main simulation parameters are listed in Table~\ref{tab:params}.

\begin{table}[h]
\centering
\caption{GUI-controllable simulation parameters.}
\label{tab:params}
\begin{tabular}{lllr}
\toprule
Parameter & Symbol & Unit & Default \\
\midrule
Pixel pitch & $p$ & \si{\micro\meter} & 500 \\
Surface resistivity & $\rho_s$ & $\Omega/\square$ & 1000 \\
Strip resistivity at electrode & $\rho_{s,\mathrm{pad}}$ & $\Omega/\square$ & 25 \\
Strip resistivity at mid & $\rho_{s,\mathrm{mid}}$ & $\Omega/\square$ & 25 \\
Strip gap at midpoint & $g_\mathrm{mid}$ & \si{\micro\meter} & 0 \\
Strip gap at electrode & $g_\mathrm{pad}$ & \si{\micro\meter} & 0 \\
Pad radius & $a$ & \si{\micro\meter} & 30 \\
Input impedance & $R_\mathrm{in}$ & \si{\ohm} & 50 \\
Signal amplitude & $A$ & a.u. & 100 \\
Noise RMS & $\sigma_n$ & a.u. & 1.0 \\
Scan steps per pixel & $n_s$ & -- & 50 \\
Events for template reco & $N_\mathrm{ev}$ & -- & 5000 \\
Trench between pixels & -- & boolean & false \\
No structure (surface only) & -- & boolean & false \\
MIP Landau charge & -- & boolean & false \\
\bottomrule
\end{tabular}
\end{table}

\subsection{Basic usage}
\label{sec:usage}

The GUI is operated in a fixed order. The first step is always to build the resistive
network for the chosen geometry and $R_\mathrm{in}$ by pressing \emph{``Compute resistance
matrix''}: this factorises the KCL matrix (Section~\ref{sec:cache}) and is a prerequisite
for every other operation -- the single-event, 2D-scan, and position-reconstruction buttons
in Fig.~\ref{fig:gui} are disabled until a matrix is available for the current parameter
set. Once the matrix exists, the position resolution can be evaluated in either of two
ways: the fast \emph{unbalance reconstruction}, which uses a closed-form combination of the
four central-electrode signals (Section~\ref{sec:reco}), or the more accurate but slower
\emph{template} method~\cite{Arcidiacono2023}, which first requires computing the charge-sharing templates
(Section~\ref{sec:templates}) and then reconstructs each event by $\chi^2$ matching against
them (Section~\ref{sec:reco}). Both the resistance matrices and the sharing templates are
cached on disk, keyed by the full set of geometry and $R_\mathrm{in}$ parameters
(Section~\ref{sec:cache}); once a given configuration has been computed, reopening the GUI
 with the same parameters loads the cached matrix and templates
directly from disk instead of repeating the resistive-network solve or the template computation.

\section{Grid and Coordinate System}
\label{sec:grid}

The simulation domain covers a $3 \times 3$ pixel area.  The grid has $n = 301$ nodes along each axis, giving a node pitch (mesh
size) of
\begin{equation}
  h = \frac{3p}{n-1} = \frac{3p}{300}.
\end{equation}
For the default pitch $p = \SI{500}{\micro\meter}$ this gives $h = \SI{5}{\micro\meter}$ per
node. The pixel pitch occupies $p/h = 100$ nodes.

Node $(i,j)$ has physical coordinates
\begin{align}
  x_{ij} &= i \cdot h, & y_{ij} &= j \cdot h,
\end{align}
with $i,j \in \{0,\ldots,300\}$. The scalar node index is $\mathrm{idx} = i \cdot n + j$.
The total number of nodes is $N_\mathrm{tot} = n^2 = 90601$.

The four electrodes in the central $2\times2$ sub-array are located at
$(r,c)\,p$ with $r,c\in\{1,2\}$ (using 0-based row/column index from 0 to 3). All 16 electrode
positions are defined analogously.

\section{Resistive Network Model}
\label{sec:network}

\subsection{Edge conductance}
\label{sec:edge}

Each pair of adjacent nodes $(i,j)$ and $(i',j')$ is connected by an edge conductance
$g_{ij\to i'j'}$ that reflects the local material properties. Three contributions are
distinguished:

\paragraph{Surface resistive layer}
The resistive surface layer has a uniform surface resistivity $\rho_s$. For a square of side $h$
the surface conductance is
\begin{equation}
  g_\mathrm{surf} = \frac{1}{R_\mathrm{surf}}, \qquad
  R_\mathrm{surf} = \rho_s.
\end{equation}
(One square contributes $\rho_s$ regardless of $h$.)

\paragraph{Resistive strip}
Along certain inter-electrode paths a  strip with lower resistivity can be present.
The strip occupies a band of width $w_\mathrm{strip} = 2\times0.025\,p = 0.05\,p$ centred on
the electrode row or column. Within this band the strip conductance is added in parallel to the
surface:
\begin{equation}
  g = g_\mathrm{surf} + g_\mathrm{strip}, \qquad
  g_\mathrm{strip} = \frac{1}{r_\mathrm{strip}(d)},
\end{equation}
where $d$ is the distance of the edge midpoint from the nearest electrode, and
$r_\mathrm{strip}$ is the local strip sheet resistance interpolated linearly between its
value at the electrode $(\rho_{s,\mathrm{pad}})$ and its value at the mid-point
$(\rho_{s,\mathrm{mid}})$:
\begin{equation}
  r_\mathrm{strip}(d) = \rho_{s,\mathrm{pad}} + (\rho_{s,\mathrm{mid}} - \rho_{s,\mathrm{pad}})
  \frac{2d}{p}.
\end{equation}

It is possible to interrupt  the strip at two locations (returning only surface conductance):
\begin{itemize}
  \item Midpoint gap $g_\mathrm{mid}$: edges within half-width $g_\mathrm{mid}/2$ of
        the inter-electrode midpoint ($d = p/2$) carry only surface conductance.
  \item Electrode gap $g_\mathrm{pad}$: edges within distance $g_\mathrm{pad}$ of any
        electrode ($d < g_\mathrm{pad}$) carry only surface conductance. This models a strip
        that ends before reaching the electrode, decoupling the strip from the electrode.
\end{itemize}

The number of squares from electrode to electrode along the strip (with no gaps) is
\begin{equation}
  N_\square = \frac{p}{w_\mathrm{strip}} = \frac{p}{0.05\,p} = 20.
\end{equation}
The total resistance from electrode to electrode is
\begin{equation}
  R_\mathrm{strip,total} = \frac{N_\square}{2}(\rho_{s,\mathrm{pad}} + \rho_{s,\mathrm{mid}})
  = 10\,(\rho_{s,\mathrm{pad}} + \rho_{s,\mathrm{mid}}).
\end{equation}
Due to the two-dimensional nature of the resistive network a correction factor of approximately
$1.7\times$ is observed with respect to the naive 1D (five-lane parallel) estimate.

\paragraph{Trench}
When the trench option is active, all strip-related conductance is suppressed and edges crossing
the inter-electrode boundaries are cut (zero conductance), except within a small radius around
each electrode. Both cardinal edges (purely horizontal or purely vertical) and diagonal edges
that cross a boundary are blocked. Only the surface contribution remains on non-trench edges. This
models an etched gap that completely interrupts current flow between pixels.

\paragraph{No structure (surface only)}
When the ``No structure'' option is active, the strip gap at midpoint is set equal to the full
pixel pitch $p$. Because the gap half-width $g_\mathrm{mid}/2 = p/2$ covers every inter-electrode
position, the strip contribution is suppressed everywhere and the surface behaves as a pure
uniform resistive surface. The strip resistance entries are disabled in the GUI to prevent
accidental modification. This mode is equivalent to a detector with no resistive strips and no
trench.

\subsection{KCL matrix assembly}
\label{sec:kcl}

The resistive network is described by the Kirchhoff Current Law (KCL) at every node. Denoting
the node voltage vector as $\mathbf{V}$ and the injected current vector as $\mathbf{B}$, the
system to be solved is
\begin{equation}
  \mathbf{M}\,\mathbf{V} = \mathbf{B},
\end{equation}
where $\mathbf{M}$ is the $N_\mathrm{tot}\times N_\mathrm{tot}$ sparse conductance (Laplacian)
matrix. Its off-diagonal entries are
\begin{equation}
  M_{\mathrm{idx},\,\mathrm{idx}'} = -g_{\mathrm{idx}\to\mathrm{idx}'},
\end{equation}
and the diagonal entries are the sum of all conductances leaving node $\mathrm{idx}$:
\begin{equation}
  M_{\mathrm{idx},\mathrm{idx}} = \sum_{\mathrm{neighbours}\,k} g_{\mathrm{idx}\to k}
  + \Delta_\mathrm{diag},
\end{equation}
where $\Delta_\mathrm{diag}$ is an additional diagonal term used to enforce boundary conditions
(see Section~\ref{sec:bc}).

\subsection{Electronics input impedance and electrode boundary conditions}
\label{sec:bc}

Each electrode is read out by a front-end amplifier whose finite input impedance $R_\mathrm{in}$
is modelled as an ordinary resistor connecting the electrode node to ground.

\paragraph{$R_\mathrm{in} = 0$} the electrode is pinned to ground ($V_k = 0$), an ideal
short.

\paragraph{$R_\mathrm{in} > 0$} the electrode is not pinned; instead a conductance $1/R_\mathrm{in}$ is added between the
electrode node and ground, in place of the ideal short:
\begin{equation}
  M_{\mathrm{el},\mathrm{el}} \mathrel{+}= \frac{1}{R_\mathrm{in}}, \qquad
  B_\mathrm{el} = 0.
\end{equation}
The electrode node is then free to rise to a non-zero potential $V_k$, and the current actually
collected by the electronics is $I_k = V_k / R_\mathrm{in}$.

Because $R_\mathrm{in}$ appears explicitly in the boundary condition at every electrode, it is
one of the network's own conductances and directly affects how charge is shared among electrodes.
Section~\ref{sec:optimisation} shows this dependence is not negligible: both the
signal fractions collected by each electrode and the position resolution vary measurably with
$R_\mathrm{in}$ over the range studied ($0$--$100\,\si{\ohm}$).

\subsection{Matrix decomposition and caching}
\label{sec:cache}

The KCL matrix depends only on the geometric and material parameters (pitch, resistivities,
gaps, trench, $R_\mathrm{in}$) and not on the injection position. It is therefore factorised
once and reused for all injection positions. The decomposition uses \texttt{TDecompSparse}
(ROOT's sparse LU factoriser).

To avoid repeated decomposition between sessions, the matrix is cached on disk in a
sub-directory \texttt{rsd\_cache/}. Two cache key formats are used:
\begin{itemize}
  \item Without trench: \texttt{PP\{p\}\_Rsh\{$\rho_s$\}\_RsP\{$\rho_{s,P}$\}\_RsM\{$\rho_{s,M}$\}\_gap\{$g_\mathrm{mid}$\}\_gapP\{$g_\mathrm{pad}$\}\_tr0\_Rin\{$R_\mathrm{in}$\}\_Rpad\{$a$\}}
  \item With trench: \texttt{PP\{p\}\_Rsh\{$\rho_s$\}\_tr1\_Rin\{$R_\mathrm{in}$\}\_Rpad\{$a$\}} (strip parameters omitted as they have no effect)
\end{itemize}
On loading, the matrix is restored using the \texttt{TMatrixDSparse} copy constructor to
avoid ROOT's incompatibility check in the assignment operator.

\section{Signal Generation and Collection}
\label{sec:signal}

\subsection{Charge injection and fluctuation model}
\label{sec:injection}

A particle crossing the detector at position $(x_0, y_0)$ is modelled as a point current source
injected at the single nearest grid node, found by rounding to the closest integer node index:
\begin{equation}
  B_\mathrm{idx} = L \cdot I_\mathrm{total}, \qquad
  \mathrm{idx} = \mathrm{round}(x_0/h)\cdot n + \mathrm{round}(y_0/h),
\end{equation}
where $I_\mathrm{total} = 1$ (normalised) and $L$ is the amplitude factor described below.

\paragraph{Charge fluctuation modes}
The GUI checkbox \emph{MIP Landau charge (laser = off)} controls the event-by-event
amplitude factor $L$. The normalised sharing fraction $f_k(x,y)$ is converted into a signal
by multiplying it by both $L$ and the GUI's \emph{Signal Amplitude [a.u.]} gain $A$ (default
$A=100$), which sets the overall scale of the collected signal independently of the
event-by-event fluctuation:

\begin{itemize}
  \item \textbf{Laser mode} ($L=1$): the injected charge is fixed, corresponding to a laser
        pulse of constant intensity. All amplitude variation comes from electronics noise only.
  \item \textbf{MIP Landau mode}: $L$ is drawn from a Landau distribution with most-probable
        value $\mathrm{MPV}=1$ and scale $\sigma_L = 0.3$, independently for each simulated
        event:
        \begin{equation}
          L \sim \mathrm{Landau}(\mathrm{MPV}=1,\,\sigma_L=0.3).
        \end{equation}
        The Gaussian noise, of fixed absolute RMS $\sigma_n$ (it does not scale with $A$ or
        $L$), is added \emph{after} scaling by both factors:
        \begin{equation}
          \hat{f}_k = A\,L\cdot f_k(x,y) + \mathcal{N}(0,\sigma_n).
        \end{equation}
\end{itemize}

\subsection{Current collection}
\label{sec:collection}

After solving $\mathbf{M}\,\mathbf{V} = \mathbf{B}$, the current collected by electrode $k$ is
computed as the net outflow from the electrode node. Two formulas apply:

\begin{itemize}
  \item $R_\mathrm{in} = 0$: $I_k = \sum_{\mathrm{neighbours}\,m} g_{k\to m}(V_k - V_m)$
        (KCL edge sum).
  \item $R_\mathrm{in} > 0$: $I_k = V_k / R_\mathrm{in}$.
\end{itemize}

The fractional sharing for pixel $k$ is $f_k = I_k / \sum_j I_j$.

\section{Transient Signal Propagation}
\label{sec:transient}

An optional transient model propagates the signal in time. Each node has a capacitance
\begin{equation}
  C_\mathrm{node} = \frac{\varepsilon_0\,\varepsilon_\mathrm{Si}\,h^2}{d_\mathrm{Si}},
\end{equation}
where $d_\mathrm{Si}$ is the silicon thickness and $h$ is the node pitch. The time step is
$\Delta t = \SI{20}{\pico\second}$. At each time step the node voltages evolve according to the
RC equation:
\begin{equation}
  C_\mathrm{node}\,\frac{dV_k}{dt} = -\sum_m g_{k\to m}(V_k - V_m) + B_k(t).
\end{equation}
This is integrated with an implicit (backward-Euler) scheme: at each step the linear system
\begin{equation}
  \left(\mathbf{G} + \frac{C_\mathrm{node}}{\Delta t}\,\mathbf{I}\right)\mathbf{V}^{n+1}
  = \frac{C_\mathrm{node}}{\Delta t}\,\mathbf{V}^{n} + \mathbf{B}^{n+1}
\end{equation}
is solved, where $\mathbf{G}$ is the conductance matrix of Section~\ref{sec:kcl}. This scheme is unconditionally stable, so the time step does not need to be smaller than the node time constant.

Signal propagation delays are also computed geometrically using a Dijkstra--Fermat
shortest-path algorithm on the resistive network, giving the time for a wavefront to travel
from the injection point to each electrode.

\section{Validity of the Quasi-Static Approximation}
\label{sec:quasistatic}

The simulation solves a static (DC) resistive network. Real detector signals from
LGAD/RSD devices are short current pulses with significant spectral content between
approximately \SI{50}{MHz} and \SI{400}{MHz}. This section quantifies the error
introduced by neglecting the finite-frequency response of the resistive sheet.

\subsection{Bulk capacitive shunting}
\label{sec:bulk_cap}

At frequency $\omega = 2\pi f$, each node of the resistive sheet is capacitively
coupled to the back-plane electrode through the silicon bulk. The capacitance per unit area is
\begin{equation}
  C_\mathrm{bulk} = \frac{\varepsilon_0\,\varepsilon_\mathrm{Si}}{d_\mathrm{Si}},
\end{equation}
where $\varepsilon_\mathrm{Si} = 11.7$ and $d_\mathrm{Si}$ is the detector thickness.
This shunt admittance modifies the 2D resistive diffusion equation into a
Helmholtz-type equation for the sheet potential:
\begin{equation}
  \nabla^2 V - \frac{j\omega\,C_\mathrm{bulk}}{\rho_s^{-1}}\,V = -\rho_s\,j_\mathrm{inj},
\end{equation}
where $j_\mathrm{inj}$ is the injected current density.
Solutions decay with a complex \emph{AC spreading length}
\begin{equation}
  \lambda_\mathrm{AC} = \frac{1}{\sqrt{j\omega\,C_\mathrm{bulk}\,\rho_s}},
  \qquad
  |\lambda_\mathrm{AC}| = \frac{1}{\sqrt{\omega\,C_\mathrm{bulk}\,\rho_s}}.
\end{equation}
The DC limit corresponds to $|\lambda_\mathrm{AC}| \to \infty$. The fractional correction
to the signal sharing at distance $p$ from the injection point is of order
$(p/|\lambda_\mathrm{AC}|)^2$.

\subsection{Numerical estimates}
\label{sec:quasistatic_numbers}

For a typical DC-RSD detector with $d_\mathrm{Si} = \SI{300}{\micro\meter}$ and
$\rho_s = \SI{1000}{\ohm}/\square$:
\begin{equation}
  C_\mathrm{bulk} = \frac{8.85\times10^{-12}\times 11.7}{300\times10^{-6}}
  \approx \SI{3.45e-7}{F/m^2}.
\end{equation}
Table~\ref{tab:lambda} lists $|\lambda_\mathrm{AC}|$ and the corresponding sharing
correction for a pixel pitch $p = \SI{450}{\micro\meter}$.

\begin{table}[h]
\centering
\caption{AC spreading length and fractional correction to signal sharing
  for $\rho_s = \SI{1000}{\ohm}/\square$, $d_\mathrm{Si} = \SI{300}{\micro\meter}$,
  $p = \SI{450}{\micro\meter}$.}
\label{tab:lambda}
\begin{tabular}{rrrr}
\toprule
$f$ & $|\lambda_\mathrm{AC}|$ & $p/|\lambda_\mathrm{AC}|$ & Correction $(p/\lambda)^2$ \\
\midrule
\SI{50}{MHz}  & \SI{9.6}{mm}  & 0.047 & $\sim 0.2\%$ \\
\SI{100}{MHz} & \SI{6.8}{mm}  & 0.066 & $\sim 0.4\%$ \\
\SI{200}{MHz} & \SI{4.8}{mm}  & 0.094 & $\sim 0.9\%$ \\
\SI{400}{MHz} & \SI{3.4}{mm}  & 0.13  & $\sim 1.7\%$ \\
\bottomrule
\end{tabular}
\end{table}

Since $|\lambda_\mathrm{AC}| \gg p$ across the full signal bandwidth, the spatial
distribution of collected charge is virtually indistinguishable from the DC case.

\subsection{Strip RC time constant}
\label{sec:strip_rc}

The resistive strips (Section~\ref{sec:edge}) have a typical resistance
$R_\mathrm{strip} \sim 500\,\Omega$ for the electrode-to-electrode path. Their self-capacitance
to the back-plane is negligible ($C_\mathrm{strip} \sim \SI{0.1}{fF}$), giving a
characteristic frequency
\begin{equation}
  f_\mathrm{3dB} = \frac{1}{2\pi\,R_\mathrm{strip}\,C_\mathrm{strip}} \sim \SI{3}{THz},
\end{equation}
which is far above the signal bandwidth. The strip impedance is therefore purely
resistive at all relevant frequencies.

\subsection{Conclusion}
\label{sec:quasistatic_conclusion}

The quasi-static (DC) approximation is valid for DC-RSD signal sharing at frequencies
up to at least \SI{400}{MHz}. The maximum fractional error in signal sharing is
$\lesssim 2\%$ over the full LGAD/RSD signal bandwidth, well below the level of
noise and other systematic effects. No frequency-dependent correction to the
resistive network model is required under typical detector parameters.

\section{Charge Sharing Templates}
\label{sec:templates}

The sharing template describes how the collected charge fraction $f_k(x,y)$ varies as a
function of the hit position $(x,y)$ within the central pixel cell. Templates are computed on
a $(n_s+1)\times(n_s+1)$ grid spanning one pixel pitch in each direction, with step size
$\Delta x = p/n_s$. The GUI parameter $n_s$ (default 50) sets the number of scan steps per
pixel; at the default, $\Delta x = 2h$ (two grid nodes per step). For each scan point the KCL
system is solved and the fractions $\{f_k\}$ for all 16 pixels are stored. The resulting
$f_k(x,y)$ maps are displayed as 2D colour histograms (\texttt{TH2D}) with $z$ range $[0,1]$.

\section{Signal Leakage}
\label{sec:leakage}

For a hit inside the central pixel cell $[p,2p]\times[p,2p]$, the signal collected by the
four central pixels (P6, P7, P10, P11) should ideally equal the total injected signal. Any
current reaching the 12 outer pixels constitutes \emph{signal leakage}:
\begin{equation}
  \mathcal{L}(x,y) = 1 - \frac{I_6 + I_7 + I_{10} + I_{11}}{I_\mathrm{total}}.
\end{equation}
The leakage map is computed over the central pixel cell by scanning at intervals of $\Delta x$
and displayed alongside the central-electrode fraction map. The surface-averaged leakage,
\begin{equation}
  \langle\mathcal{L}\rangle = \frac{1}{N_\mathrm{bins}}\sum_{\mathrm{bins}} \mathcal{L}(x,y),
\end{equation}
is printed to the terminal after the scan. The two maps share the same scan and are computed
simultaneously without additional solver calls.

Leakage is largest near the pixel boundaries, where the signal is shared between pixels, and
is enhanced when a resistive strip at low resistivity provides a direct current path to outer
electrodes.

\section{Position Reconstruction}
\label{sec:reco}

\subsection{Template-based $\chi^2$ minimisation}
\label{sec:chi2}

Given a measured set of signal fractions $\{\hat{f}_k\}$ (with Gaussian noise of RMS
$\sigma_n$ added), the reconstructed position $(\hat{x}, \hat{y})$ is found by minimising
\begin{equation}
  \chi^2(x,y) = \sum_{k \in \{6,7,10,11\}} [\hat{f}_k - f_k(x,y)]^2,
\end{equation}
where the sum runs over the \textbf{four central electrodes} only (P6, P7, P10, P11), since the
templates are built exclusively for these pixels.
The minimisation scans the $(n_s+1)\times(n_s+1)$ template grid to find the coarse minimum, then applies
parabolic interpolation in both $x$ and $y$ independently to obtain sub-grid precision:
\begin{equation}
  x^* = x_m - \frac{1}{2}\,\frac{\chi^2_{m+1} - \chi^2_{m-1}}{\chi^2_{m+1} - 2\chi^2_m + \chi^2_{m-1}}\,\Delta x.
\end{equation}
Parabolic interpolation removes the scan-direction bias that would otherwise produce an
asymmetric $x$-residual distribution.

\subsection{Unbalance reconstruction and distortion grid}
\label{sec:noiseless}

A reconstructed position is computed analytically:
\begin{align}
  x_\mathrm{nl} &= p\cdot\frac{3}{2} + \frac{(f_{10}+f_{11})-(f_6+f_7)}{f_6+f_7+f_{10}+f_{11}}\,\frac{p}{2}, \label{eq:unbalance_x}\\
  y_\mathrm{nl} &= p\cdot\frac{3}{2} + \frac{(f_7+f_{11})-(f_6+f_{10})}{f_6+f_7+f_{10}+f_{11}}\,\frac{p}{2}, \label{eq:unbalance_y}
\end{align}
using the unbalance formula with noiseless amplitudes. The noiseless reconstruction is used
to populate the \emph{distortion grid} ($x_\mathrm{true}$ vs $x_\mathrm{nl}$) as a scatter
plot, which shows the systematic nonlinearity of the unbalance estimator without contamination
from noise. A separate noisy realisation is used to build the distortion map displayed as a
2D histogram.

\subsection{Resolution map}
\label{sec:resmap}

The position resolution is evaluated by generating $N_\mathrm{ev}$ random hit positions uniform
in $[p, 2p] \times [p, 2p]$ (the central pixel cell). For each hit the reconstruction error is
$\epsilon_x = \hat{x} - x_\mathrm{true}$. The RMS resolution map is computed on a
$(n_s+1)\times(n_s+1)$ grid, where each bin accumulates the squared errors:
\begin{equation}
  \sigma_x(x,y) = \sqrt{\frac{1}{N_\mathrm{bin}}\sum_{i\in\mathrm{bin}} \epsilon_{x,i}^2}.
\end{equation}
This definition includes both the statistical spread and any residual bias; it is equivalent
to the root-mean-square position error. The $z$ scale is fixed at $[0, \SI{20}{\micro\meter}]$.

\subsection{Resolution fits}
\label{sec:fits}

The global $x$-residual distribution $\Delta x = \hat{x} - x_\mathrm{true}$ is fitted with
two models displayed in separate panels:

\paragraph{Gaussian fit}
A standard Gaussian $\mathcal{N}(\mu,\sigma)$ is fitted to extract the core resolution
$\sigma$.

\paragraph{Student's $t$ fit}
A Student's $t$ distribution with location $\mu$, scale $\sigma_\mathrm{scale}$, and
$\nu$ degrees of freedom is fitted over the range $[-\SI{100}{\micro\meter},\,+\SI{100}{\micro\meter}]$:
\begin{equation}
  f(x;\,\mu,\sigma_\mathrm{scale},\nu) =
  \frac{1}{\sigma_\mathrm{scale}}\,
  t\!\left(\frac{x-\mu}{\sigma_\mathrm{scale}};\,\nu\right),
\end{equation}
where $t(z;\nu)$ is the standard Student's $t$ density implemented via
\texttt{TMath::Student}. For $\nu > 2$ the true standard deviation of the distribution is
\begin{equation}
  \sigma_\mathrm{RMS} = \sigma_\mathrm{scale}\sqrt{\frac{\nu}{\nu-2}}.
\end{equation}
The $t$ distribution accounts for heavy tails arising from ambiguous template matches near
pixel boundaries and from Landau amplitude fluctuations (MIP mode). As $\nu\to\infty$ it
converges to the Gaussian. Fit parameters
($\sigma_\mathrm{scale}$, $\nu$, $\sigma_\mathrm{RMS}$) are reported on the terminal.

\section{Inter-electrode Resistance Measurement}
\label{sec:resistance}

The effective resistance between two electrodes is measured using a two-terminal
method. For each pair of electrodes $(P_a, P_b)$ to be measured:

\begin{enumerate}
  \item Build the free KCL matrix with all electrode nodes floating (no boundary conditions applied to electrodes). This matrix is independent of $R_\mathrm{in}$.
  \item Add a weak reference conductance $G_\mathrm{ref} = 10^{-9}$ to node 0 to remove the
        singular null mode:
        \begin{equation}
          M_{0,0} \mathrel{+}= G_\mathrm{ref}.
        \end{equation}
  \item Set the right-hand side $\mathbf{B}$ with $B(P_a) = +1$ (current source) and
        $B(P_b) = -1$ (current sink).
  \item Solve $\mathbf{M}\,\mathbf{V} = \mathbf{B}$.
  \item The inter-electrode resistance is $R_{ab} = V(P_a) - V(P_b)$.
\end{enumerate}

This approach leaves all other electrodes floating so that their presence does not short-circuit the
measurement path. Two measurements are performed:

\begin{itemize}
  \item $R_\mathrm{side}$: pixel $P_{11}$ (row 2, col 2) to $P_7$ (row 1, col 2) --- side neighbours.
  \item $R_\mathrm{diag}$: pixel $P_{11}$ (row 2, col 2) to $P_6$ (row 1, col 1) --- diagonal neighbours.
\end{itemize}

In a pure 2D resistive sheet the resistance between two electrodes scales logarithmically with
their separation, so the ratio $R_\mathrm{diag}/R_\mathrm{side}$ is
expected to be slightly above unity:
\begin{equation}
  \frac{R_\mathrm{diag}}{R_\mathrm{side}} \approx \frac{\ln(\sqrt{2}\,p/a)}{\ln(p/a)},
\end{equation}
where $a$ is the effective electrode radius. Numerically this gives a ratio of approximately
$1.13$--$1.15$, consistent with simulation results.

\section{Summary of Simulation Workflow}
\label{sec:workflow}

\begin{enumerate}
  \item \textbf{Grid setup}: generate the $301\times301$ node grid for the chosen pitch $p$.
  \item \textbf{Matrix build}: assemble the KCL conductance matrix \textbf{M} using
        \texttt{GetEdgeConductance()} for each of the four neighbour directions at every node.
        Apply electrode boundary conditions according to $R_\mathrm{in}$.
  \item \textbf{Decomposition}: factorise \textbf{M} with \texttt{TDecompSparse}.
        Cache the matrix and decomposition on disk using a parameter-encoded file name.
  \item \textbf{Template scan}: for each of the $(n_s+1)\times(n_s+1)$ scan positions within the central
        pixel cell, inject current, solve, collect currents, store $\{f_k(x,y)\}$.
  \item \textbf{Signal leakage scan}: for each scan position in the central pixel cell, compute
        the fraction of signal reaching outer pixels; report the surface-averaged leakage.
  \item \textbf{Noiseless reconstruction}: for each scan point compute $x_\mathrm{nl}$, $y_\mathrm{nl}$
        using the unbalance formula with noiseless amplitudes; populate the linearity scatter plot.
  \item \textbf{Event loop}: generate $N_\mathrm{ev}$ random hits; optionally draw amplitude factor
        $L$ from a Landau distribution (MIP mode) or set $L=1$ (laser mode); add Gaussian noise;
        reconstruct positions via $\chi^2$ minimisation with parabolic interpolation.
  \item \textbf{Resolution map}: bin reconstruction errors into $(n_s+1)\times(n_s+1)$ map, compute
        RMS per bin.
  \item \textbf{Resolution fits}: fit the global $\Delta x$ distribution over $[-100,+100]\,\mu$m
        with a Gaussian and a Student's $t$ (via \texttt{TMath::Student});
        report $\sigma$, $\sigma_\mathrm{scale}$, $\nu$, $\sigma_\mathrm{RMS}$.
  \item \textbf{Linearity}: display $x_\mathrm{true}$ vs $x_\mathrm{reco}$ scatter with diagonal reference.
  \item \textbf{Resistance}: Kelvin measurement between selected electrode pairs (independent of $R_\mathrm{in}$).
\end{enumerate}

\section{Validation against test-beam data}
\label{sec:validation}

The simulation is validated against test-beam data from FBK DC-RSD sensors with trench
isolation. For a representative operating point (pixel pitch $p = 500$~\si{\micro\meter},
bias voltage 240~V, signal MPV $\approx 91$~mV, noise RMS $= 2.5$~mV, surface resistivity
$\rho_s = 800$~$\Omega/\square$, trench enabled, constant term $k_2 = 0.03$).
The simulation is run with signal amplitude $A = 91$ and noise $\sigma_n = 2.5$, instead of the default values of Table~\ref{tab:params}.
The residual distribution $\Delta x = \hat{x} - x_\mathrm{true}$ is fitted with a Student's
$t$ distribution both in data and simulation. The data give $\sigma_\mathrm{scale} = 22.4$~\si{\micro\meter}
and $\nu = 3.13$, while the simulation yields $\sigma_\mathrm{scale}^{\mathrm{sim}} = 22.0$~\si{\micro\meter}
and $\nu^{\mathrm{sim}} = 3.74$, in good agreement.

\begin{figure}[h]
\centering
  \includegraphics[width=.95\linewidth]{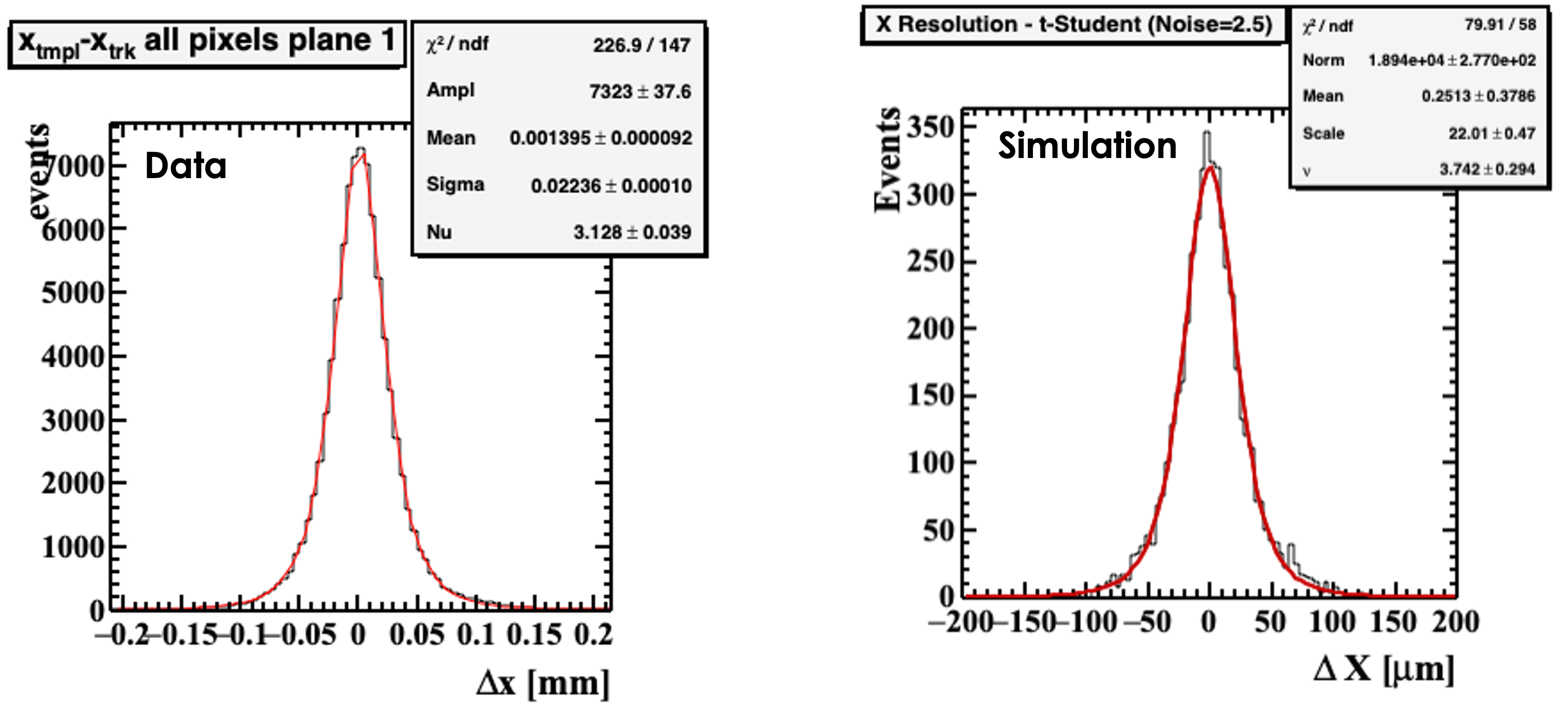}
\caption{Residual distributions $\Delta x$ for data (left) and simulation (right) for a
500~\si{\micro\meter} DC-RSD at 240~V bias. Both are fitted with a Student's $t$ distribution.}
\label{fig:residuals}
\end{figure}

\begin{figure}[h]
\centering
 \includegraphics[width=.85\linewidth]{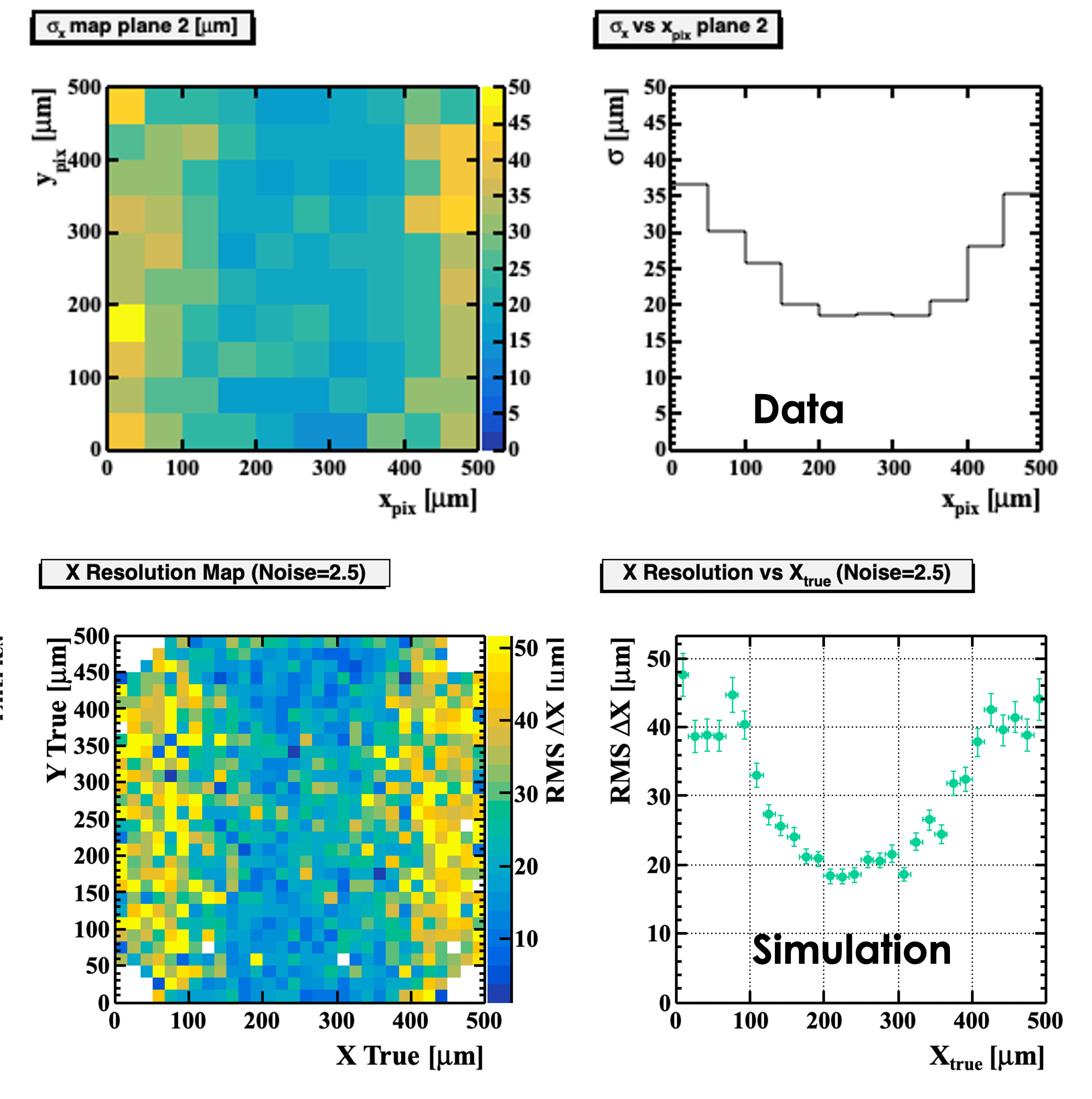}
\caption{Resolution map $\sigma_x(x,y)$ (left: data, right: simulation) and 1D projection
vs $x_\mathrm{true}$ for a 500~\si{\micro\meter} DC-RSD.}
\label{fig:resmap}
\end{figure}

The agreement extends to the spatial distribution of the resolution. Figure~\ref{fig:resmap}
shows the 2D resolution map $\sigma_x(x,y)$ and the 1D projection $\sigma_x$ vs
$x_\mathrm{true}$, comparing test-beam data and simulation. The characteristic non-uniformity
— low resolution near the electrode centres, higher resolution near the pixel edges — is
reproduced quantitatively.

The signal-sharing templates — the 2D maps of charge fraction $f_k(x,y)$ for each of the
four central electrodes — also show excellent agreement between data and simulation, confirming
that the resistive network model correctly captures the lateral charge spread on the sensor
surface.

\begin{figure}[h]
\centering
 \includegraphics[width=.85\linewidth]{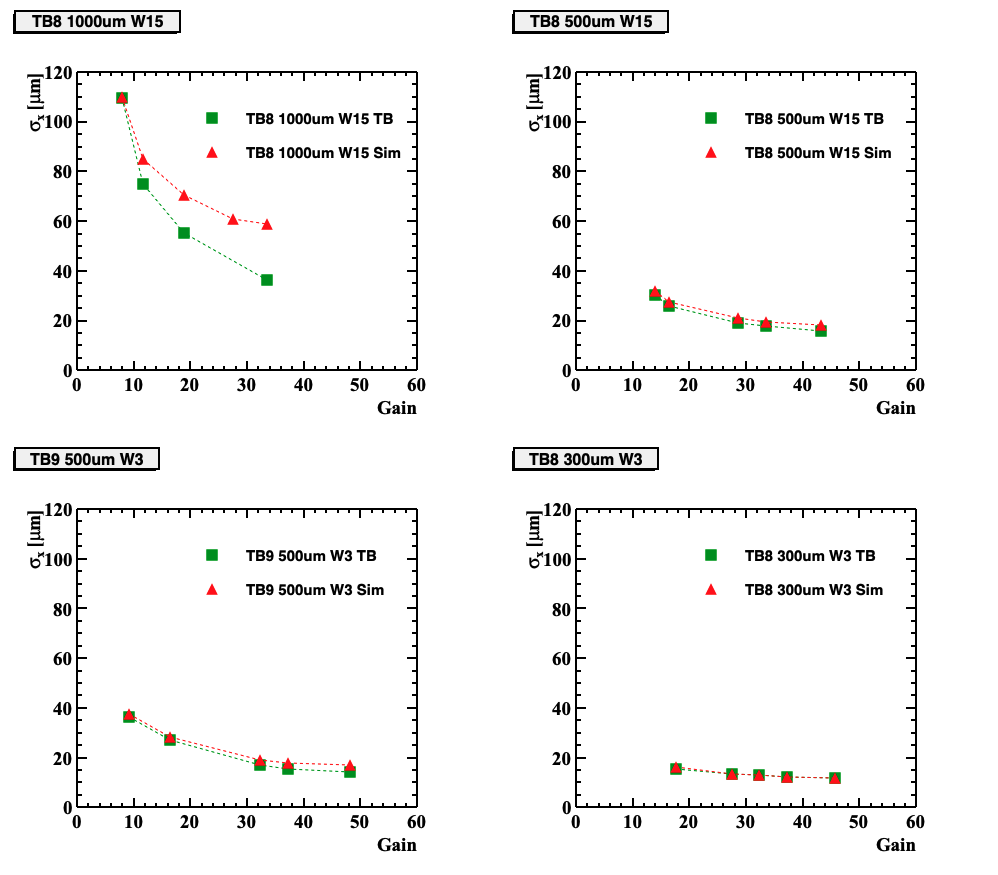}
\caption{Position resolution $\sigma_x$ vs gain for four DC-RSD sensor configurations,
reconstructed with the template method. Filled squares: template built from test-beam data
(tracker-truth position). Filled triangles: template from the simulation.}
\label{fig:resvsgain}
\end{figure}

A broader validation exploits the fact that a charge-sharing template can be built directly
from test-beam data: the external tracker provides an independent, precise measurement of
where each particle actually crossed the sensor, so the average electrode-by-electrode charge fraction
can be binned as a function of this known, true position exactly as is done for the
simulated templates of Section~\ref{sec:templates}. Because it is built from the true hit
position rather than from a model, this \emph{test-beam template} captures the real
charge-sharing behaviour of the sensor -- including any effects not implemented in the
simulation -- and the position resolution obtained by reconstructing test-beam events
against it is therefore the best resolution achievable with the template method for that
sensor. Reconstructing the same test-beam events against the \emph{simulated} template
instead, and comparing the two resolutions, is thus a direct test of the simulation: if the
simulated template reproduces the real charge-sharing shape, the two reconstructions must
agree, since the reconstruction algorithm and the input events are identical and only the
template differs. Figure~\ref{fig:resvsgain} shows this comparison as a function of gain for
four sensor configurations (500~\si{\micro\meter}~W15 from TB8, 500~\si{\micro\meter}~W3
from TB9, 1000~\si{\micro\meter}~W15 from TB8, and 300~\si{\micro\meter}~W3 from TB8). The
very good agreement between the test-beam-template and simulated-template resolutions across
all configurations and gain values confirms that the simulated templates closely reproduce
the real ones, and hence that the simulator correctly captures the signal split on the
sensor surface.

\section{Optimisation of inter-electrode structures}
\label{sec:optimisation}

Having validated the simulator, it is used to explore three inter-electrode containment
strategies for DC-RSD sensors:
\begin{itemize}
  \item \textbf{No structure}: pure resistive surface, no inter-electrode element.
  \item \textbf{Trench}: physical etch that interrupts current flow between pixels (modelled
        as zero conductance on cross-boundary edges).
  \item \textbf{Resistive strip}: a low-resistivity strip running between electrodes,
        added in parallel to the surface layer.
\end{itemize}
Unless otherwise stated, every comparison in this section uses a surface resistivity
$\rho_s = 1000\,\Omega/\square$ and an electrode radius $a = 30\,\si{\micro\meter}$, both in line
with the values realised in the FBK DC-RSD production used for the test-beam validation of
Section~\ref{sec:validation}, while the electronics input impedance $R_\mathrm{in}$ is
scanned over $0$--$100\,\si{\ohm}$. All three parameters are held identical across the five
geometries at every scan point, so that only the inter-electrode structure itself differs
between the configurations being compared.
Each option is evaluated against three figures of merit: (i) position resolution and its
spatial uniformity, (ii) inter-electrode resistance, and (iii) signal leakage to outer
pixels.

\subsection{Position resolution: unbalance reconstruction}
\label{sec:res_unbalance}

\begin{figure}[h]
\centering
 \includegraphics[width=.95\linewidth]{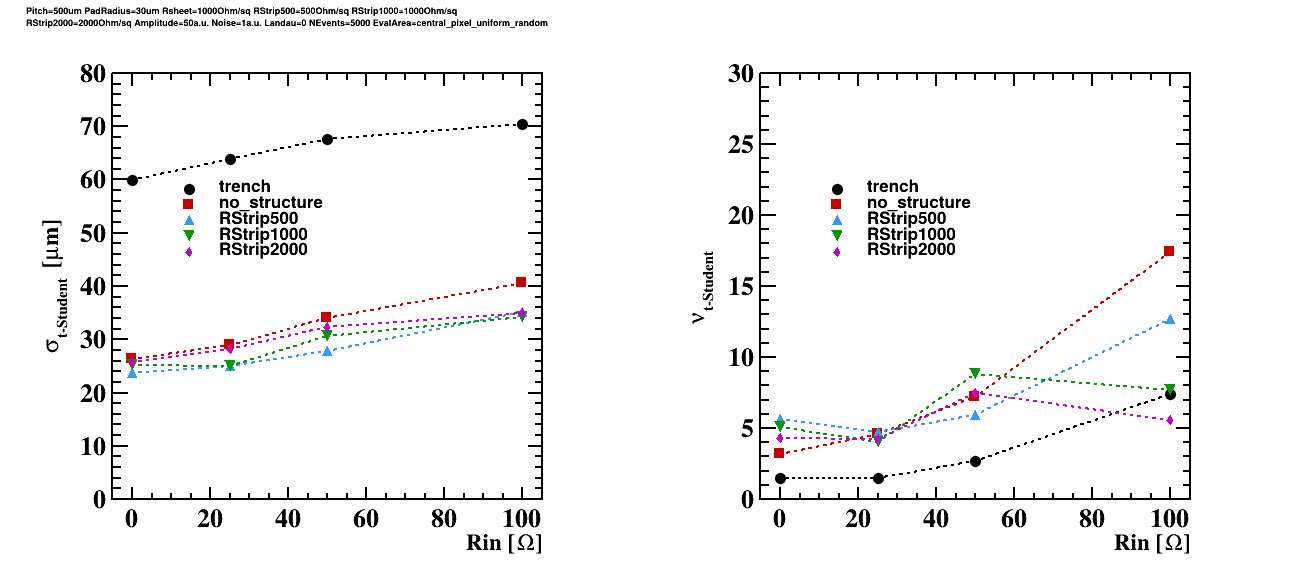}
\caption{Unbalance-reconstruction resolution $\sigma_\mathrm{scale}$ (left) and tail
parameter $\nu$ (right) vs $R_\mathrm{in}$, for trench, no structure, and three resistive
strips ($500,1000,2000\,\Omega$). The trench case is worse than any other configuration at
every $R_\mathrm{in}$.}
\label{fig:unbalance}
\end{figure}

A first, model-independent estimate of the achievable resolution is obtained with the
unbalance estimator of Eqs.~(\ref{eq:unbalance_x})--(\ref{eq:unbalance_y}), evaluated with noise added directly to the collected
amplitudes rather than through the full template $\chi^2$ fit. Figure~\ref{fig:unbalance}
shows the Student's $t$ fit parameters ($\sigma_\mathrm{scale}$ and $\nu$) obtained this way
for the trench, no-structure, and three resistive-strip configurations
($R_\mathrm{strip} = 500, 1000, 2000\,\Omega$), as a function of $R_\mathrm{in}$.
The trench case is $30$--$60\%$ worse than any of the strip configurations or the bare
surface at every value of $R_\mathrm{in}$, and its $\nu$ parameter is systematically lower,
indicating heavier residual tails. The resistive-strip configurations track the bare surface
closely, with the lowest-resistivity strip ($500\,\Omega$) consistently the best performer.

\subsection{Charge-sharing templates}
\label{sec:templates_compare}

\begin{figure}[h]
\centering
 \includegraphics[width=.95\linewidth]{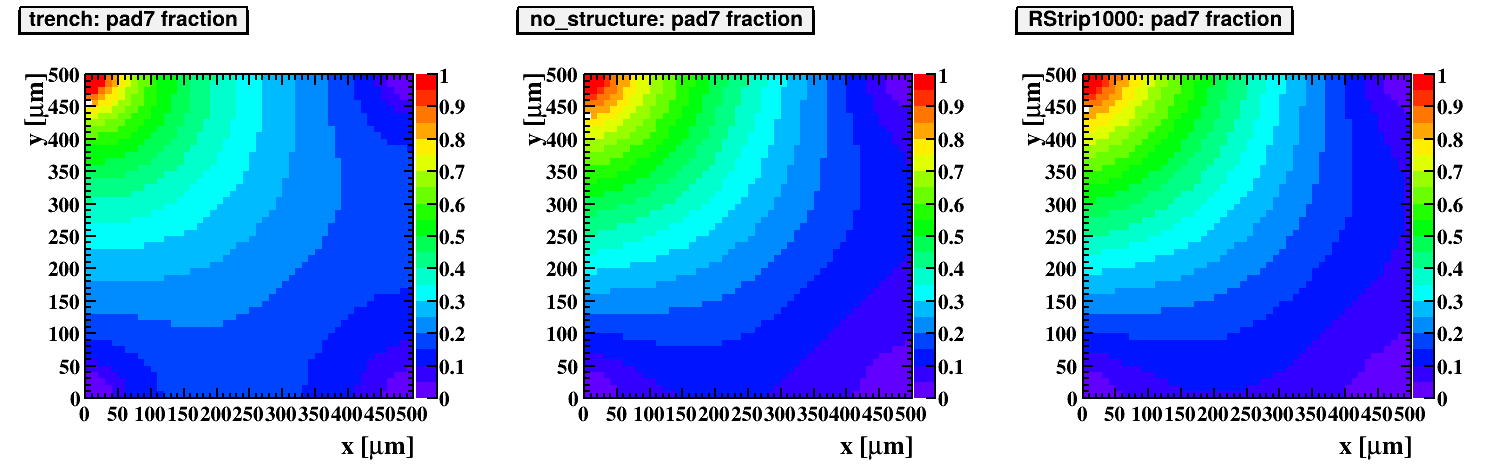}
\caption{Charge-sharing fraction of the top-left electrode as a function of position, for
trench, no structure, and a $1\,$k\si{\ohm} resistive strip.}
\label{fig:template_pad7}
\end{figure}

To understand the origin of this difference, the charge-sharing templates
themselves are examined. Figure~\ref{fig:template_pad7} shows the fraction of signal collected by a
single electrode (the top-left corner of the central pixel) as a function of position, for
the trench, no-structure, and $1\,$k\si{\ohm}-strip cases. The three maps are qualitatively
similar in shape and dynamic range: what matters for position reconstruction is not the map
itself but \emph{how quickly the fraction changes when moving across the pixel}, which is
examined next.

\subsection{Position sensitivity: signal in the two left electrodes}
\label{sec:leftsum}

\begin{figure}[h]
\centering
 \includegraphics[width=.95\linewidth]{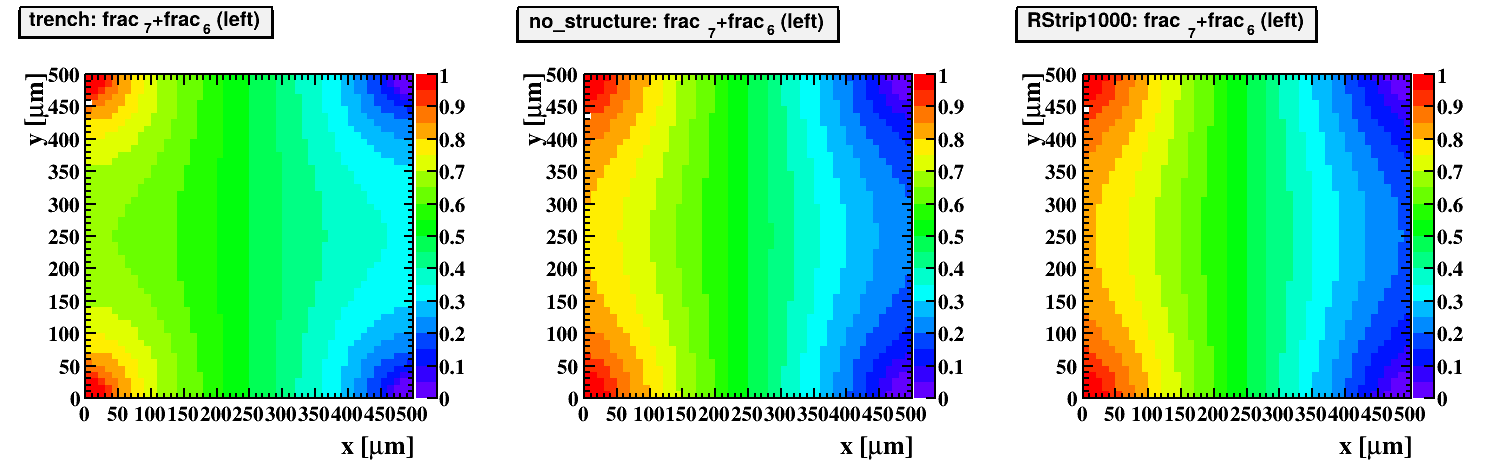}
\caption{Combined signal fraction of the two left electrodes, $f_6+f_7$, vs position, for
trench, no structure, and a $1\,$k\si{\ohm} resistive strip. The contours are nearly
vertical (a clean function of $x$ alone) for no structure and the resistive strip, but bow
inward for the trench.}
\label{fig:leftsum}
\end{figure}

The natural variable for reconstructing the $x$ coordinate is the combined signal fraction
in the two left electrodes, $f_6 + f_7$ (the same quantity that enters the numerator of the
unbalance formula, Eq.~(\ref{eq:unbalance_x})). Figure~\ref{fig:leftsum} shows this sum as a function of
position for the three configurations. For no structure and for the resistive strip the
iso-contours are nearly vertical, i.e.\ $f_6+f_7$ depends almost exclusively on $x$ at every
height $y$ -- a clean, effectively one-dimensional response. For the trench the contours
bow markedly inward at mid-height, so that the sum retains a significant, spurious
dependence on $y$: the trench response is two-dimensional in a way the other two
structures are not.

\subsection{Position sensitivity: gradient analysis}
\label{sec:gradient}

\begin{figure}[h]
\centering
 \includegraphics[width=.95\linewidth]{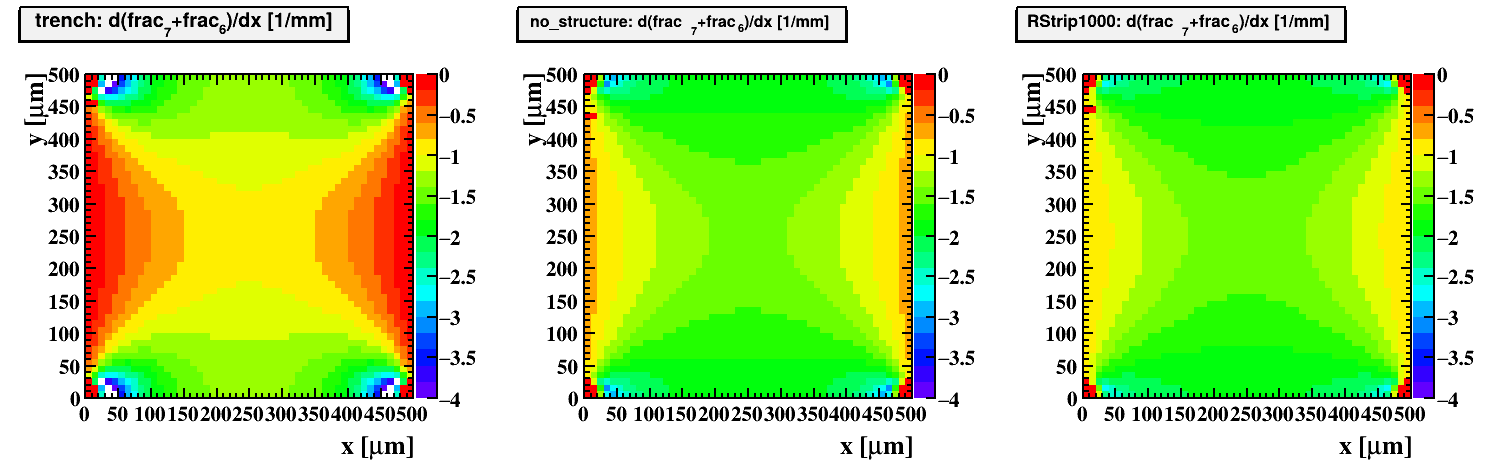}
\caption{Gradient $\partial(f_6+f_7)/\partial x$ vs position, for trench, no structure, and
a $1\,$k\si{\ohm} resistive strip ($z$ axis capped at $-4\,\mathrm{mm}^{-1}$ to make the
interior structure visible; the near-electrode singularity, identical for all three geometries by
construction, is saturated). The trench has both the weakest gradient and the largest
weak-gradient area.}
\label{fig:leftsum_grad}
\end{figure}

The position resolution scales inversely with the local gradient of the reconstruction
variable: $\sigma_x \propto \sigma_n / |\partial(f_6+f_7)/\partial x|$ for a fixed noise
$\sigma_n$. Figure~\ref{fig:leftsum_grad} shows $\partial(f_6+f_7)/\partial x$, computed by
finite differences on the templates of Fig.~\ref{fig:leftsum}. The trench gradient is
markedly weaker than that of the other two structures over most of the pixel area, and the
region of strong gradient is confined to narrow bands next to the shared corner electrodes.
A quantitative comparison at two representative points (pixel centre and near the pixel
edge, at mid-height) makes this explicit:
\begin{center}
\begin{tabular}{lcc}
\toprule
 & Trench & No structure \\
\midrule
Gradient at pixel centre & $-0.93\,\mathrm{mm}^{-1}$ & $-1.44\,\mathrm{mm}^{-1}$ \\
Gradient near pixel edge & $-0.16\,\mathrm{mm}^{-1}$ & $-0.82\,\mathrm{mm}^{-1}$ \\
\bottomrule
\end{tabular}
\end{center}
At the centre the trench gradient is already $35\%$ weaker than the bare surface, consistent
with the classical resistive-division result that sensitivity is maximal at the point
equidistant from all electrodes. Moving toward the edge, however, the bare-surface gradient
stays strong ($-0.82\,\mathrm{mm}^{-1}$) while the trench gradient collapses by a further
factor of $\sim6$ to $-0.16\,\mathrm{mm}^{-1}$, nearly flat. Because the position resolution
is sampled uniformly over the whole pixel area, and the trench's weak-gradient region
covers most of that area, this collapse away from the centre -- not a uniform reduction in
sensitivity -- is the dominant reason the trench under-performs in Fig.~\ref{fig:unbalance}
and in the strip-resistance scan below.

\subsection{Position resolution vs $R_\mathrm{in}$}
\label{sec:res_rin}

\begin{figure}[h]
\centering
 \includegraphics[width=.95\linewidth]{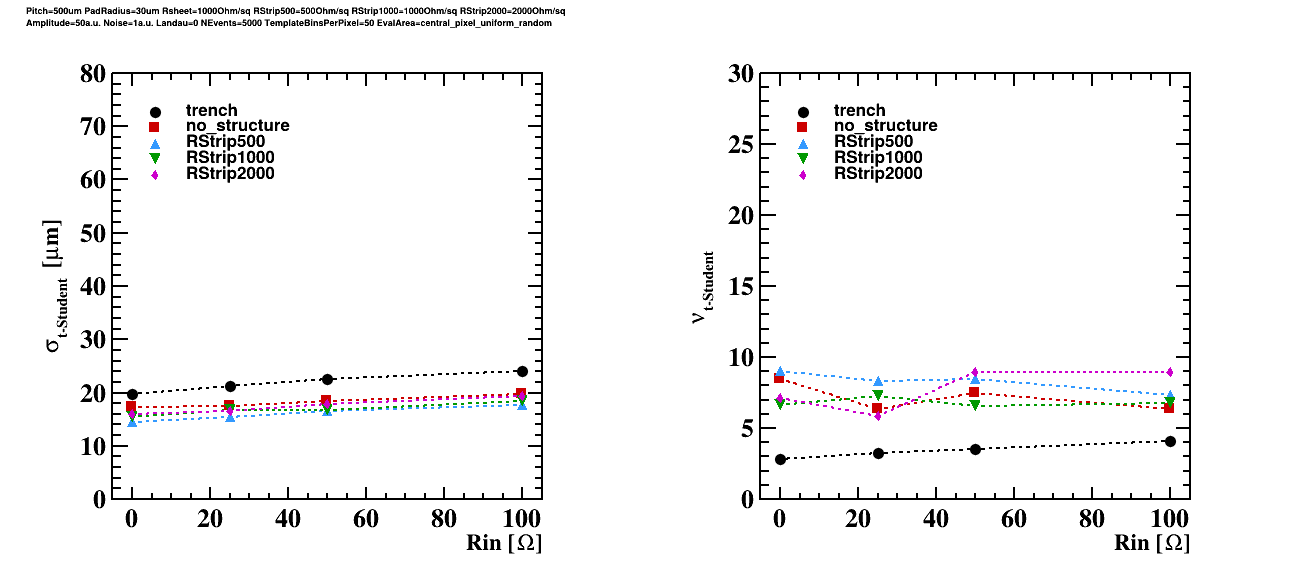}
\caption{Template-based position resolution $\sigma_\mathrm{scale}$ (left) and tail
parameter $\nu$ (right) vs $R_\mathrm{in}$, for trench, no structure, and three resistive
strips ($500,1000,2000\,\Omega$).}
\label{fig:templatefit_rin}
\end{figure}

Having identified the physical mechanism, it is now characterised quantitatively as a
function of the electronics input impedance $R_\mathrm{in}$, using the full
template-based $\chi^2$ reconstruction of Section~\ref{sec:reco} for all five
configurations. Figure~\ref{fig:templatefit_rin} shows the Student's $t$ fit parameters
($\sigma_\mathrm{scale}$ and $\nu$) vs $R_\mathrm{in}$.

At every $R_\mathrm{in}$ the ranking is $\mathrm{RStrip500} < \mathrm{RStrip1000} <
\mathrm{RStrip2000} < \text{no structure} < \text{trench}$, matching the mechanistic
explanation of Sections~\ref{sec:res_unbalance}--\ref{sec:gradient}. At $R_\mathrm{in}=0$
the trench resolution ($19.7\,\si{\micro\meter}$) is $36\%$ worse than the best strip
($14.5\,\si{\micro\meter}$); at $R_\mathrm{in}=100\,\si{\ohm}$ it is $24.0\,\si{\micro\meter}$,
still $35\%$ worse than the best strip ($17.7\,\si{\micro\meter}$) -- the trench penalty is
essentially independent of $R_\mathrm{in}$. The $\nu$ parameter for the trench
($2.8$--$4.1$) is systematically far below that of every other configuration
($6.3$--$9.0$), confirming the heavier residual tails expected from the gradient collapse
identified in Section~\ref{sec:gradient}.

\subsection{Signal leakage vs $R_\mathrm{in}$}
\label{sec:leakage_rin}

\begin{figure}[h]
\centering
 \includegraphics[width=.48\linewidth]{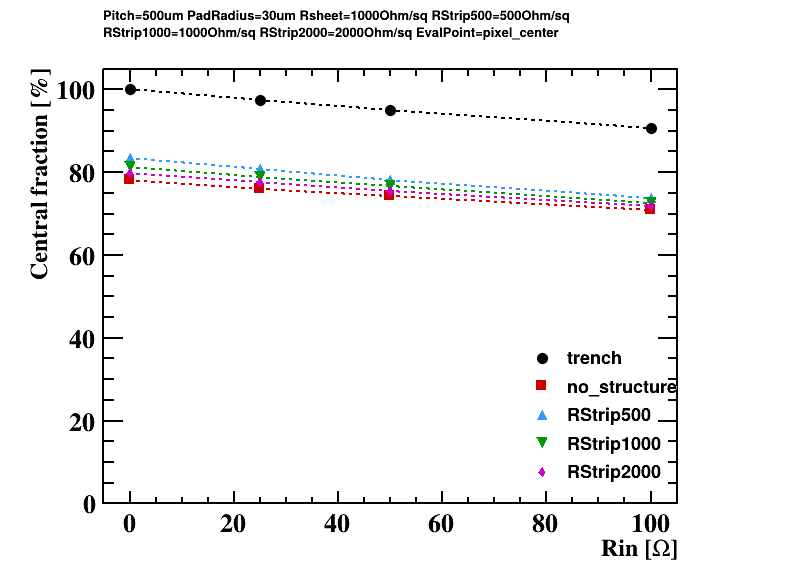}\hfill
 \includegraphics[width=.48\linewidth]{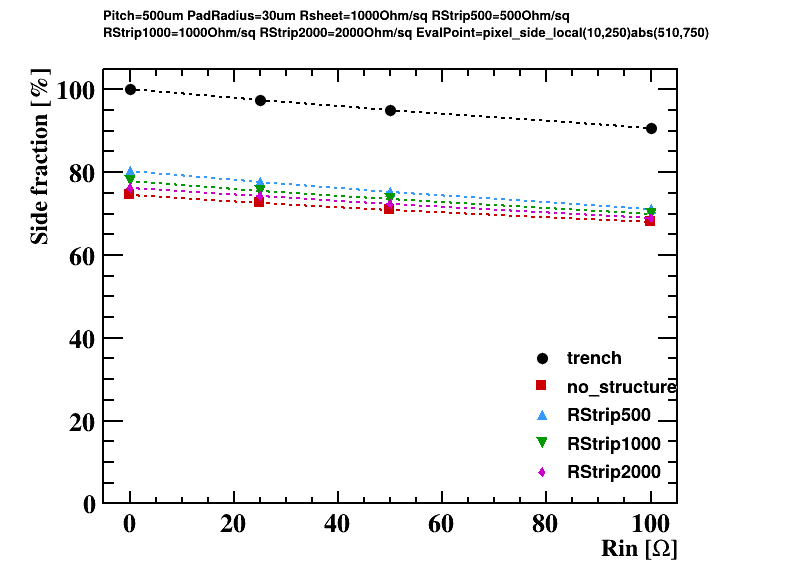}
\caption{Fraction of injected charge collected by the four central electrodes vs
$R_\mathrm{in}$, evaluated at the pixel centre (left) and near a pixel side (right), for all
five configurations. The trench values are identical in the two panels.}
\label{fig:fraction_rin}
\end{figure}

The signal fraction collected by the four central electrodes (the complement of the
leakage to the outer pixels) is shown in Fig.~\ref{fig:fraction_rin}, both at the pixel
centre and near a pixel side ($10\,\si{\micro\meter}$ from the edge). The two figures reveal
a trade-off that is the counterpart of the resolution result above: the trench, precisely
because it electrically isolates each pixel, has leakage less than 10\%
($100\%\to90.7\%$ central fraction from $R_\mathrm{in}=0$ to $100\,\si{\ohm}$, i.e.\ up to
$9\%$ leakage only through the $R_\mathrm{in}$ path itself), while the resistive-strip and
no-structure designs leak $16$--$30\%$ of the signal to neighbouring pixels over the same
range -- yet it is exactly this leaked charge that carries the positional information
exploited in Sections~\ref{sec:leftsum}--\ref{sec:gradient}. Notably, the trench values are
\emph{identical} at the pixel centre and near the pixel side, confirming numerically that
its charge-sharing response does not depend on position within the pixel, whereas the other
configurations show a  lower containment near the pixel side than at the centre
(e.g.\ no structure: $78.0\%\to74.5\%$ at $R_\mathrm{in}=0$), consistent with a
charge-sharing gradient that stays non-zero all the way to the pixel boundary.

\subsection{Inter-electrode resistance vs $R_\mathrm{in}$}
\label{sec:interR_rin}

The inter-electrode resistance is measured using the two-terminal method described in
Section~\ref{sec:resistance}, with an a posteriori correction $R_\mathrm{eff} = R_\mathrm{geometric} +
2R_\mathrm{in}$, which adds the input impedance of the two probed electrodes in series (the
purely geometric network resistance discussed in Section~\ref{sec:bc} is itself independent
of $R_\mathrm{in}$, but $R_\mathrm{eff}$, what an external ohmmeter would actually measure,
is not). The trench configuration could not be included in this scan because its
free-floating resistance matrix (all electrodes floating except the two Kelvin probes) proved
numerically unstable in the sparse solver; only its $R_\mathrm{in}=0$
reference values are therefore quoted, $R_\mathrm{side}^\mathrm{trench} = 2.2$~k\si{\ohm} and
$R_\mathrm{diag}^\mathrm{trench} = 2.5$~k\si{\ohm}, without a full $R_\mathrm{in}$ scan.

\begin{figure}[h]
\centering
 \includegraphics[width=.95\linewidth]{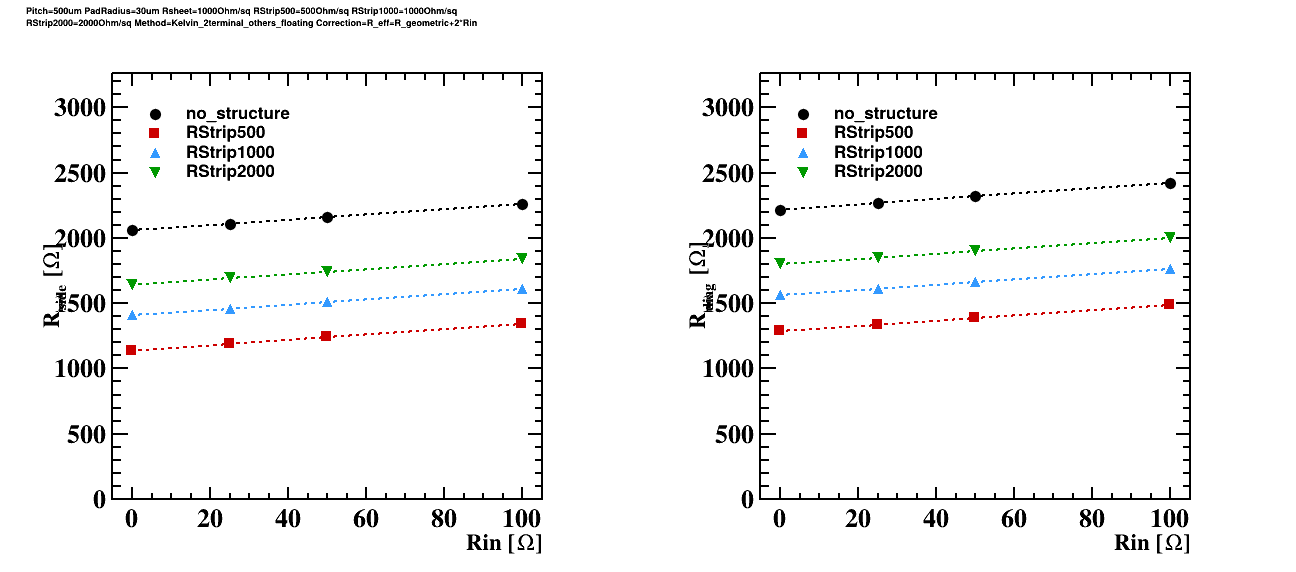}
\caption{Inter-electrode resistance $R_\mathrm{eff}$ vs $R_\mathrm{in}$ for side- (left) and
diagonal-neighbour (right) pixel pairs, for no structure and three resistive strips. The
trench could not be scanned (solver instability) but its $R_\mathrm{in}=0$ values,
$2.2/2.5$~k\si{\ohm}, exceed all configurations shown here at every $R_\mathrm{in}$.}
\label{fig:interpad_rin}
\end{figure}

As expected from the correction term, every configuration's resistance grows linearly with
$R_\mathrm{in}$, by exactly $2R_\mathrm{in}$ ($100\,\si{\ohm}$ increase over the full scan
range). At fixed $R_\mathrm{in}$ the resistance increases monotonically with
$R_\mathrm{strip}$, from $\approx1.14/1.28$~k\si{\ohm} (side/diagonal) for RStrip500, the
lowest of all configurations tested ($\sim50\%$ of the trench value), up to
$\approx1.64/1.80$~k\si{\ohm} for RStrip2000. The no-structure case, with no additional
conduction path at all, reaches the highest resistance among the four
($\approx2.06/2.22$~k\si{\ohm}), only $\sim7\%$ below the trench reference: any resistive
strip, however high its own sheet resistance, sits electrically in parallel with the bulk
sheet and can therefore only lower the net inter-electrode resistance relative to the
no-structure case, never raise it. This makes it fundamentally difficult to reach very high
inter-electrode resistance with a resistive-strip design; only the trench, which removes
the parallel path altogether, reaches values around $2.2$--$2.5$~k\si{\ohm}.

\subsection{Trench advantages beyond position resolution}
\label{sec:trench_other}

The results above show that resistive strips give substantially better position resolution
than trench, and reach a higher inter-electrode resistance only with difficulty. Two further
considerations, not addressed by the simulation, nevertheless favour trench isolation in
specific operating conditions.

\emph{Timing resolution at low charge}. Because trench confines essentially all of
the collected charge to the struck pixel at every hit position (Section~\ref{sec:leakage_rin}),
the largest-signal electrode of that pixel receives a bigger fraction of the total charge
than the corresponding electrode of a resistive-strip or bare-surface design, where the same
total charge is shared more evenly, and over a wider area, among the four central electrodes.
Since the jitter contribution to timing resolution scales inversely with the collected
signal amplitude, a larger single-electrode signal directly improves that electrode's own
time measurement. This advantage is most relevant at low total charge, where the timing
jitter term dominates: trench then reduces the amount of charge \emph{sharing} exactly where
resistive designs would otherwise waste part of the already-scarce charge over multiple
electrodes.

\emph{Disambiguation of two nearly simultaneous hits in adjacent pixels}. Two
neighbouring pixels share either one or two electrodes. With trench isolation no charge can cross a trench boundary other than through
the electrode metal itself, so if two particles strike two adjacent pixels at (nearly) the same
time, only the shared electrodes collect contributions from both signals while the other
electrodes carry the signal of one particle only, exactly as in a single-particle event.
Each hit can therefore still be reconstructed independently combining independent and common electrodes~\cite{Arcidiacono2026}.
In a resistive-strip or bare-surface design charge is not confined to the struck
pixel and spreads measurably beyond the pixels immediately bordering it (Section~\ref{sec:leakage_rin}), so a two-particle event in adjacent pixels can leave a mixed
signal on more than the one geometrically shared electrode, making the two hits harder to
disentangle.

\section{Conclusions}
\label{sec:conclusions}

This work has addressed the question of which inter-electrode structure -- an isolating
trench, a purely resistive surface, or a low-resistivity strip -- gives the best position
resolution in a DC-RSD sensor. To this end, \texttt{DCRSD\_SurfaceSimulator} was built, a
fast 2D resistive network simulator for DC-coupled Resistive Silicon Detectors, which was
validated against test-beam measurements. The key results are:
\begin{itemize}
  \item The simulator correctly reproduces test-beam measurements from FBK DC-RSD sensors
        with 300, 500, and 1000~\si{\micro\meter} pixel pitch, including the value of the
        position resolution, its spatial uniformity, and the signal-sharing templates.
  \item Resistive strips give a position resolution up to $\sim$25\% better than an
        isolating trench, essentially independent of $R_\mathrm{in}$, while the bare
        surface (no structure at all) performs almost as well as the strips. Trench
        under-performs because its charge-sharing gradient -- the quantity that sets the
        resolution -- collapses to near zero away from the pixel centre, whereas the
        gradient of the other two designs stays strong across most of the pixel.
  \item The choice of surface structure is therefore a trade-off, not a single optimum: a
        resistive strip or a bare surface should be preferred where position resolution is
        the priority.  The same charge confinement that limits trench's
        position resolution also gives it two further advantages not addressed by the
        simulation: better timing resolution at low charge and better
        disambiguation of two nearly simultaneous hits in adjacent pixels.
  \item The simulator is publicly available and can be used to optimise the surface geometry
        of future DC-RSD productions.
\end{itemize}

\section*{Code and Documentation Availability}
\label{sec:availability}

The source code \texttt{DCRSD\_SurfaceSimulator.c} and this technical note are publicly
available on GitHub:
\begin{center}
  \url{https://github.com/cartiglia/DCRSD_SS}
\end{center}
To download the code, clone the repository with:
\begin{lstlisting}[language=bash]
git clone https://github.com/cartiglia/DCRSD_SS.git
\end{lstlisting}
The simulator requires ROOT 6 (\url{https://root.cern}) and is run directly as a ROOT macro:
\begin{lstlisting}[language=bash]
root DCRSD_SurfaceSimulator.c
\end{lstlisting}

\section*{Acknowledgements}
The support of the following Funding Agencies is gratefully acknowledged: INFN CSN5
through the STARS research project; European Union's Horizon Europe Research and
Innovation programme Grant Agreement No.\ 101057511 (EUROLABS). The fruitful discussions
with the DRD3 collaboration, CERN, are also acknowledged. The measurements leading to these
results have been performed at the test-beam facility at DESY Hamburg (Germany), a member
of the Helmholtz Association (HGF), whose support is gratefully acknowledged.



\begin{thebibliography}{9}

\bibitem{Menzio2022}
  L.~Menzio \emph{et al.},
  \emph{DC-coupled resistive silicon detectors for 4D tracking},
  Nucl.\ Instrum.\ Methods A \textbf{1041} (2022) 167374 [arXiv:2204.07226].

\bibitem{Fondacci2025TCAD}
  A.~Fondacci \emph{et al.},
  \emph{Design and optimisation of radiation resistant AC- and DC-coupled resistive LGADs},
  2025 JINST \textbf{20} C06016.

\bibitem{Cartiglia2026LGAD}
  N.~Cartiglia \emph{et al.},
  \emph{Why (and how) LGAD work: ionization, space charge, and gain saturation},
  Nucl.\ Instrum.\ Methods A \textbf{1090} (2026) 171674.

\bibitem{Cenna2015}
  F.~Cenna \emph{et al.},
  \emph{Weightfield2: A fast simulator for silicon and diamond solid state detector},
  Nucl.\ Instrum.\ Methods A \textbf{796} (2015) 149--153.

\bibitem{DESYTestbeam}
  M.~Stanitzki, S.~Ackermann, R.~Diener and N.~Meyners,
  \emph{The DESY II Test Beam Facility -- Current Status and Future Plans},
  2024 IEEE Nuclear Science Symposium (NSS), Medical Imaging Conference (MIC) and Room
  Temperature Semiconductor Detector Conference (RTSD), Tampa, FL, USA, 2024, pp.\ 1--1,
  doi:10.1109/NSS/MIC/RTSD57108.2024.10657589.

\bibitem{CentisVignali2025}
  M.~Centis Vignali \emph{et al.},
  \emph{Development and wafer-level characterization of the first production of DC-RSD
  sensors at FBK},
  2025 JINST \textbf{20} C07037.

\bibitem{Arcidiacono2026}
  R.~Arcidiacono,
  \emph{Status and perspectives of DC-coupled Resistive Silicon Detectors},
  talk given at Vertex 2026, \url{https://indico.cern.ch/event/1618157/overview},
  proceedings in preparation.
  
  \bibitem{Cassese2026}
  A.~Cassese,
  \emph{Design review and performance of the FBK DC-RSD production},
  talk given at IWORID 2026,  proceedings in preparation.
  
\bibitem{Arcidiacono2023}
  R.~Arcidiacono \emph{et al.},
  \emph{High precision 4D tracking with large pixels using thin resistive silicon
  detectors},
  Nucl.\ Instrum.\ Methods A \textbf{1057} (2023) 168671.

\end{thebibliography}
\end{document}